\documentclass[paper,twocolumn,showpacs,superscriptaddress,preprintnumbers,floats,aps,pra,10pt]{revtex4-1}
\usepackage{graphicx}
\usepackage{dcolumn}
\usepackage{amsmath}
\usepackage{amsfonts}
\usepackage{amssymb}
\usepackage{color}
\usepackage{float}
\usepackage{pstricks}
\usepackage{natbib}
\usepackage{array}
\usepackage{multirow}
\usepackage{ulem}
\providecommand{\U}[1]{\protect\rule{.1in}{.1in}}

\DeclareMathOperator\arctanh{arctanh}

\begin{document}

\title{A non-singular early-time viscous cosmological model}
\author{Norman Cruz}
\altaffiliation{norman.cruz@usach.cl}
\affiliation{Departamento de F\'isica, Universidad de Santiago de Chile, \\
Avenida Ecuador 3493, Santiago, Chile}
\affiliation{Center for Interdisciplinary Research in Astrophysics and Space Exploration (CIRAS),Universidad de Santiago de Chile, Avenida Libertador Bernardo O'Higgins 3363, Estación Central, Chile}
\author{Esteban Gonz\'alez}
\altaffiliation{esteban.gonzalez@ucn.cl}
\affiliation{Departamento de F\'{i}sica, Universidad Cat\'{o}lica del Norte, Avenida Angamos 0610, Casilla 1280, Antofagasta, Chile}
\author{Jose Jovel}
\altaffiliation{jose.jovel@usach.cl}
\affiliation{Departamento de F\'isica, Universidad de Santiago de Chile, \\
Avenida Ecuador 3493, Santiago, Chile}

\date{\today}

\begin{abstract}
\begin{center}
\textbf{{Abstract:}}
\end{center}
In this paper, we study the thermodynamical and mathematical consistencies for a non-singular early-time viscous cosmological model known as soft-Big Bang, which was previously found in [N. Cruz, E. Gonz\'alez, and J. Jovel, Phys. Rev. D \textbf{105}, 024047 (2022)]. This model represents a flat homogeneous and isotropic universe filled with a dissipative radiation fluid and a cosmological constant $\Lambda$, which is small but not negligible, in the framework of Eckart's theory. In particular, we discuss the capability of the solution in the fulfillment of the three following conditions: (i) the near equilibrium condition, which is assumed in Eckart's theory of non-perfect fluids, (ii) the mathematical stability of the solution under small perturbations, and (iii) the positiveness of the entropy production. We have found that this viscous model can describe the radiation domination era of the $\Lambda$CDM model and, at the same time, fulfill the three conditions mentioned by the fulfillment of a single constraint on the bulk viscous coefficient $\xi_{0}$, finding also that this non-singular model has a positive energy density in the infinity past which is infinity hotter with a constant entropy.
\vspace{0.5cm}
\end{abstract}
 \maketitle


\section{Introduction}
Currently, it is well known that the standard cosmological model, namely $\Lambda$CDM, is the most simple and successful model in order to describe the actual background cosmological data \cite{Planck2018,BAO2017,WMAP2013,OHD2021}. In this model, the Universe is described at the current time by the dark sector, roughly classified by 70$\%$ of dark energy (DE), modeled by a positive cosmological constant (CC) $\Lambda$, which is responsible for the recent acceleration in the Universe expansion \cite{sPerlmutter}; and 30$\%$ of cold dark matter (CDM), modeled by a pressureless fluid, and which is responsible for the structure formation in the Universe \cite{LSCDM}. Also, this model considers a flat Friedman-Lemaître-Robertson-Walker (FLRW) metric due to the homogeneity and isotropy of the Universe at large scales \cite{Planck2018,BAO2017} (the so-called \textit{cosmological principle} \cite{librocaro}). A well known characteristic of the $\Lambda$CDM model is the presence of an initial singularity in the early-time known as ``Big-Bang" \cite{GLBigBang}, behavior that is not a surprise considering that many solutions of the Einstein field equations drive to different singularities \cite{Hawking:1973uf}.


There are some cosmological models with the attractive quality of avoiding the initial Big-Bang singularity, in which the universe does not have a beginning of time, and therefore, avoids a quantum regime for space-time. For example, a regular scenario called ``soft Big-Bang" is discussed in \cite{primerarticulo,Softbang1,SoftBang2,Rosen,BigBounce}, describing universes that start in an eternal physical past time that comes from a static universe. Other models correspond to the so-called emergent and bouncing universes \cite{universoemergente,Ellis:2002we,Starobinsky,otrobouncing,Generalized,linking,Mukherjee:2005zt}, with the particularity of avoiding also future singularities like the ``Big-Rip'' singularities. In this sense, and going further than the $\Lambda$CDM paradigm in which all the matter components of the Universe are described by perfect fluids, a universe without Big-Bang singularity was described in a solution found by Murphy in \cite{Big.Bang}, where viscosity is present in the matter component. Following this line, the presence of regular universes for imperfect fluid was discussed in Ref.\cite{modelosinbigbang}, being an interesting point of view considering the fact that the role of viscosity in the early-time seems to be significant \cite{creaciondemateriabulk,Cosmologycreation,creaciondemateriabulk1,creaciondemateriabulk2,earlyandlateviscous,stabilidadprofe,inflacionOditsov}. For example, and considering that in a homogeneous and isotropic universe only the bulk viscosity is important, in \cite{creaciondemateriabulk,Cosmologycreation,creaciondemateriabulk1,creaciondemateriabulk2} it was discussed the relation between particle creation and bulk viscosity in the early-time evolution. Also in \cite{earlyandlateviscous,stabilidadprofe,inflacionOditsov} it was shown that bulk viscosity has an important role in the inflation process since the effect of bulk viscosity is to produce a negative pressure that drives an acceleration in the universe expansion \cite{paperprofeAccelerated,Bulk1,Bulk4,Bulk5,Bulk6,Bulk8,Almada1} (more reference about the importance of bulk viscosity in the evolution of the Universe can be seen in \cite{Oditsov1,Oditsov2,Almadaconstraints,Bulk2,Bulk3,Bulk7,Exact,Testing}). On the other hand, it is important to mention that when we attempting to explore the unknown physics of the very early Universe, i.e., near a singularity or in attempting to avoid a singularity in models such as bouncing universe (or before a non-singular bounce occurs) \cite{BigBounce,universoemergente,Generalized}, the anisotropic effects become significant, and the shear viscosity play a important role \cite{Ganguly:2021pke,PhysRevD.88.103521,Ganguly:2019llh}.

In order to describe a viscous cosmological model, a theory of non-perfect fluid out of equilibrium is needed. Such theory was first developed by Eckart in \cite{Eckart}, which is a first-order theory, with a similar theory proposed by Landau an Lifshitz in \cite{LandauandLifshitz}. Later, it was shown in \cite{NocausalEckart,Muller} that the Eckart's theory is non-causal. To avoid this non-physical issue, a causal theory was proposed by Israel and Stewart (IS) in \cite{I.S.1,I.S.2}, which is capable to reduce to Eckart's theory when the relaxation time for the bulk viscous effects is negligible \cite{Dissipativecosmology}. Accordingly, Eckart's theory is widely considered in the literature as a first approximation to describe viscous cosmology since the IS theory presents a much greater mathematical difficulty \cite{AlmadaEckartbulktest,bulkyaceleracion,chinos,articuloprofephantom,Brevik,primerarticulo,segundoarticulo,Testing2}. It is important to mention that, both theories are constructed assuming the so-called \textit{near equilibrium condition}, which was widely discussed by Maartens in the context of dissipative inflation \cite{Dissipativecosmology}. This condition refers to the physical assumption that the theories are close to thermodynamic equilibrium, and is translated into that the viscous pressure $\Pi$ has to be lower than the equilibrium pressure $p$ of the dissipative fluid, i.e.,
\begin{equation}\label{nearequilibrium}
l=\left|\frac{\Pi}{p}\right|\ll 1.
\end{equation}
According to Maartens, an accelerated expansion due only to the negativeness of the viscous pressure $\Pi$ implies a direct contradiction with the near equilibrium condition \eqref{nearequilibrium}. In this sense, the authors in \cite{Analysing,articulobueno,segundoarticulo} propose that if a positive CC is considered in these theories, then the near equilibrium condition could be fulfilled in some regimes. As a matter of fact, this viscous pressure $\Pi$ is proportional to the bulk viscosity coefficient $\xi$, which depends, particularly, on the temperature and pressure of the viscous fluid \cite{librocaro}. Therefore, a natural election for the bulk viscous coefficient for the dissipative fluid is a dependency proportional to the power of their energy density, an election that has also been explored in the literature in the context of singularities \cite{Big.Bang,rho1,Brevik,linking,tiposdeBigRip,clasificacionBigRipdetallada,primerarticulo}.

For a universe dominated only by perfect fluids, there is no entropy production since the thermodynamics of these fluids are reversible. Nevertheless, for cosmologies with non-perfect fluids, where irreversible process exists, the entropy production has to be positive for all the cosmic evolution \cite{Tamayo,fullsolution,Exact,entropiaprofe,Mar2}. Despite the fact that the Eckart and IS theories are constructed assuming a positive entropy production, some solutions can violate this condition for a certain range of values of their respective free parameters \cite{Tamayo} or it could grow to infinity in a future Big-Rip singularity \cite{segundoarticulo}, and hence, the study of the positiveness of the entropy production can lead to constraints on the free parameters of the model. In this sense, a complementary study is the study of the mathematical stability of the solutions, which can lead to more constraints on the free parameters of the model, being used both constraints for further comparison with the constraints obtained from the fulfillment for the near equilibrium condition in order to obtain a suitable solution compatible with this two thermodynamic conditions plus the mathematical one.

The near equilibrium condition, mathematical stability, and entropy production for viscous fluids have been widely investigated in the literature. For example, the near equilibrium condition was studied in \cite{fullsolution} for the IS theory where the authors consider a gravitational constant $G$ and  CC $\Lambda$ that vary over time; while in \cite{DissipativeBoltzmann} it was studied in both Eckart and IS theories for the case of a dissipative Boltzmann gas. The mathematical stability was studied, in particular, in \cite{estabilidadmatematica1} in the IS theory for a universe filled with one viscous fluid, whose bulk viscosity obeys a power law in the energy density of the dissipative fluid and without the inclusion of a CC; while in \cite{stabilidadprofe} it was studied in the de Sitter phase of cosmic expansion when the source of the gravitational field is a viscous fluid. The entropy production was studied in \cite{Tamayo} in the Eckart and IS theories where the dissipative effects are present in the DE component; while in \cite{Exact} the authors study the entropy production in the IS theory considering only a dissipative matter component, discussing also the thermodynamics properties of their solutions (more excellent references for entropy production in viscous fluids can be found in \cite{producciondeentropia,Titus,produccionentropia2,produccionentropia3,produccionentropia4,produccionentropia5}).

The aim of this paper is to study the near equilibrium condition, the mathematical stability, and the positiveness of the entropy production of a non-singular early-time viscous cosmological model within Eckart's theory in a flat FLRW metric, given by an analytical solution found in \cite{primerarticulo}. The model itself is dominated by only two fluids: (i) a dissipative radiation component, and (ii) a DE modeled by the CC, which is small but not negligible; and was obtained assuming a bulk viscosity of the form $\xi=\xi_{0}\rho$, where $\xi_{0}>0$ is the bulk viscous coefficient \cite{librocaro}. The main motivation to study this solution, obtained for a particular case of bulk viscosity, is due to the fact that this solution behaves very close to the $\Lambda$CDM model during all the cosmic time when $\xi_{0}\to 0$, with a de Sitter type expansion in the very late-times regardless the type of the viscous fluid and the value of the bulk viscous constant; but without an initial singularity, with a behavior known as ``soft-Big Bang'' \cite{Softbang1,SoftBang2}. In addition, it was recently found in \cite{segundoarticulo} that this particular case can describe the combined  SNe Ia + OHD data in the same way as the $\Lambda$CDM model when a dissipative Warm DM (WDM) is considered instead of radiation.
We study the near equilibrium condition, the mathematical stability, and the entropy production of this solution in order to find the constraints that these criteria impose on the model's free parameters, focusing on the possibility to have a range of them satisfying all of these conditions in early-times, with the motivation to obtain a suitable non-singular early-time viscous solution from the physical and mathematical point of view.

The outline of this paper is as follows: In section \ref{seccion2} we resume the solution found in \cite{primerarticulo} that represents the model under study. In section \ref{tresproblemas} we present general results about the near equilibrium condition, the mathematical stability, and the entropy production. In Sec. \ref{exactearly} we study the exact solution to early-times, where in Sec. \ref{nearequilibrionearly} we study the near equilibrium condition, in Sec. \ref{estabilidadearlyexacta} we study the mathematical stability in the Hubble parameter, and in Sec. \ref{seccion5C} we study the entropy production. Finally, in Sec. \ref{seccionfinal} we present some conclusions and final discussions. $8\pi G=c=1$ units will be used in this work.

\section{Exact analytical solution in Eckart's theory with CC}\label{seccion2}

We will dedicate this section to summarizing a de Sitter-like solution and an analytical solution found in \cite{primerarticulo}, for a flat FLRW universe composed by a dissipative fluid ruled by a barotropic equation of state (EoS) of the form $p=(\gamma-1)\rho$, where $p$ and $\rho$ are the equilibrium pressure and energy density of the dissipative fluid, respectively, with a bulk viscosity that obeys the power law $\xi=\xi_{0}\rho^{m}$, where $\xi_{0}>0$ from the second law of thermodynamics \cite{librocaro}; and a DE component given by the CC $\Lambda$. In the framework of Eckart's theory, the field equations are given by \cite{Eckart,primerarticulo,segundoarticulo}
\begin{equation}\label{tt}
H^2=\Big(\frac{\dot{a}}{a}\Big)^{2} =\frac{\rho}{3}+\frac{\Lambda}{3},
\end{equation}
\begin{equation}\label{rr1}
\frac{\ddot{a}}{a}=\dot{H}+H^2=-\frac{1}{6} \left(\rho+3P_{eff}\right)+\frac{\Lambda}{3},
\end{equation}
plus the conservation equation
\begin{equation}\label{ConsEq1}
\dot{\rho}+3H(\rho+P_{eff})=0,
\end{equation}
where $P_{eff}$ and $\Pi$ are the effective pressure and the bulk viscous pressure of Eckart's theory, respectively, which are defined as
\begin{equation}\label{Peff}
P_{eff}=p+\Pi,
\end{equation}
\begin{equation}\label{Pi}
\Pi=-3H\xi.
\end{equation}
In the above equations, $H$ and $a$ are the Hubble parameter and the scale factor, respectively, and ``dot" accounts for derivative with respect to cosmic time $t$. From Eqs. \eqref{tt}-\eqref{Pi} a first-order differential equation for the Hubble parameter is obtained and is given by

\begin{equation}\label{Hpunto}
2\dot{H}+3\gamma H^{2}-3\xi_{0}H(3H^{2}-\Lambda)^{m}-\Lambda\gamma=0.
\end{equation}

This last expression was studied in Ref. \cite{primerarticulo} for the particular cases of $m=0$ and $m=1$, in the context of late and early-time singularities, with the consideration of a positive and negative CC. Additionally, these results were compared with the $\Lambda$CDM model and, to do this, the differential equation \eqref{Hpunto} is solved for the case of $\xi_{0}=0$, and taking  the initial conditions  $H(t=0)=H_{0}$ and $a(t=0)=1$, leading to
\begin{equation}\label{Hestandar}
H(t)=\frac{H_{0}\sqrt{\Omega_{\Lambda_{0}}}\left(\left(\sqrt{\Omega_{\Lambda_{0}} }+1\right) e^{3 \gamma  H_{0} t \sqrt{\Omega_{\Lambda_{0}} }}-\sqrt{\Omega_{\Lambda_{0}} }+1\right)}{\left(\sqrt{\Omega_{\Lambda_{0}} }+1\right) e^{3\gamma H_{0} t \sqrt{\Omega_{\Lambda_{0}} }}+\sqrt{\Omega_{\Lambda_{0}} }-1}, \end{equation}
\begin{equation}\label{aestantdar}
    a(t)=\left(\cosh\left(\frac{3\gamma\sqrt{\Omega_{\Lambda_{0}}}H_{0}t}{2}\right)+\frac{\sinh\left(\frac{3\gamma\sqrt{\Omega_{\Lambda_{0}}}H_{0}t}{2}\right)}{\sqrt{\Omega_{\Lambda_{0}}}}\right)^{\frac{2}{3\gamma}},
\end{equation}
where $\Omega_{\Lambda_{0}}=\Lambda/(3H^{2}_{0})$. As we can see from this solution, there are   no future singularities since  Eq.(\ref{Hestandar}) tends asymptotically at very late times ($t\to\infty$) to $H=\sqrt{\Lambda/3}$ which is known as the de Sitter solution \cite{deSitter} where the matter density is null, according to Eq. \eqref{tt}, but it suffers of a initial singularity in the early-time known as ``Big-Bang", as can be see in Fig. \ref{Tm1pasado}. 

The bulk viscous solution of particular interest to us is the case of $m=1$ for Eq. \eqref{Hpunto} with a positive CC. For this case, a de Sitter-like solution ($\dot{H}=0$) is found and is given by
\begin{eqnarray}
 \label{H1deSitter} 
E_{dS}&=&\frac{\gamma}{\Omega_{\xi_{0}}},
\end{eqnarray}
where $E(t)=H(t)/H_{0}$. A exact analytical solution ($\dot{H}\neq 0$) is also obtained with the border condition $H(t=0)=H_{0}$, and in a dimensionless form, is given by

\begin{eqnarray}\label{Tm1}
&\tau=\frac{\Omega_{\xi_{0}} \sqrt{\Omega_{\Lambda_{0}}} \log \left(\frac{(1-\Omega_{\Lambda_{0}} ) (\gamma - E\Omega_{\xi_{0}} )^2}{(E^2-\Omega_{\Lambda_{0}} (\gamma - \Omega_{\xi_{0}} )^2}\right)}{{3 \sqrt{\Omega_{\Lambda_{0}} } \left(\gamma^2-\Omega^{2}_{\xi} \Omega_{\Lambda_{0}} \right)}} \nonumber \\
& +\frac{\gamma  \log \left(\frac{\left(\sqrt{\Omega_{\Lambda_{0}} }-1\right) \left(\sqrt{\Omega_{\Lambda_{0}} }+E\right)}{\left(\sqrt{\Omega_{\Lambda_{0}} }+1\right) \left(\sqrt{\Omega_{\Lambda_{0}} }-E\right)}\right)}{{3 \sqrt{\Omega_{\Lambda_{0}} } \left(\gamma^2-\Omega^{2}_{\xi} \Omega_{\Lambda_{0}} \right)}},
\end{eqnarray}
where $\tau=tH_{0}$ and $\Omega_{\xi_{0}}=3\xi_{0}H_{0}$. The above solution is an implicit relation of $E(\tau)$.

The main characteristics of the de Sitter-like solution given by Eq. \eqref{H1deSitter} and the exact solution given by Eq. \eqref{Tm1} are:
\begin{itemize}
    \item[(i)]  De Sitter-like solution: This solution was previously found in \cite{Big.Bang} for the case of a null CC. This solution is related only to the dissipative processes, and the barotropic index. In addition, it was previously found in \cite{primerarticulo} that the condition $\rho>0$ restricts the value of the CC by 
\begin{equation}\label{condicionLcero}
\Omega_{\Lambda_{0}}<{\gamma^{2}}/{\Omega^{2}_{\xi_{0}}},    
\end{equation}
which means that the value of the CC has an upper limit that can grow with small values of $\Omega_{\xi_{0}}$ until the limit in with $\Omega_{\Lambda_{0}}=1$, i.e., $\Omega_{\xi_{0}}\leq\gamma$, in which the condition \eqref{condicionLcero} is always fulfilled. Also, this asymptotic behavior has a constant energy density, which can be seen from Eq. \eqref{tt}, and is given in a dimensionless form as
\begin{equation}\label{rhodeSitter1}
    \Omega_{m}=\frac{\gamma^{2}}{\Omega^{2}_{\xi_{0}}}-\Omega_{\Lambda_{0}}.
\end{equation}
Contrary to the usual de Sitter solution, this de Sitter-like solution doesn't have a null energy density, except when $\Omega_{\Lambda_{0}}=\gamma^{2}/\Omega_{\xi_{0}}^{2}$. 

\item[(ii)] Late-time behavior of the exact solution: Following the reference \cite{primerarticulo}, the exact solution given by Eq. \eqref{Tm1} tends asymptotically to late-times to the usual de Sitter solution $E=\sqrt{\Omega_{\Lambda_{0}}}$, as long as the following condition is fulfilled
\begin{equation}\label{condicionBigRip}
    \Omega_{\xi_{0}}<\gamma,
\end{equation}
which also leads to a universe with a behavior very similar to the $\Lambda$CDM model, which coincides when $\Omega_{\xi_{0}}\to 0$. In addition, in \cite{segundoarticulo} it was recently shown that, if the condition \eqref{condicionBigRip} is hold, then the exact viscous solution given by Eq. \eqref{Tm1} can describe the combined  SNe Ia + OHD data in the same way as the $\Lambda$CDM model when a dissipative WDM ($\gamma\gtrsim 1$) is considered (instead of a dissipative radiation component).

\begin{figure}[H]
\includegraphics[width=0.5\textwidth]{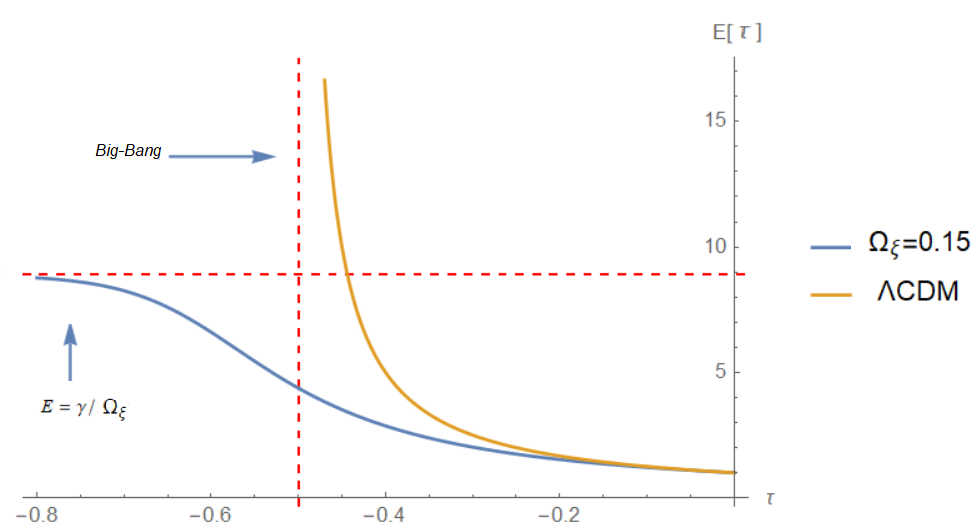}
\caption{Numerical behavior of $E(\tau)$, obtained from Eq. \eqref{Tm1} at early-times, for $\Omega_{\Lambda_{0}}=10^{-6}$, $\Omega_{\xi_{0}}=0.15$ and $\gamma=4/3$. For a comparison, we also plotted the $\Lambda$CDM model obtained from Eq. \eqref{Hestandar}.}
\label{Tm1pasado}
\end{figure}

\item[(iii)] Early-time behavior of the exact solution: As it was discussed in \cite{primerarticulo}, if the condition \eqref{condicionBigRip}
is fulfilled, then the solution represents a universe without singularity towards the past, as can be seen in  Fig. \ref{Tm1pasado}, where numerically was found the behavior of $E$ as a function of $\tau$ from Eq. \eqref{Tm1}, with the values for the free parameters of $\gamma=4/3$ and $\Omega_{\xi_{0}}=0.15$. For these values, we are considered an early-time radiation domination era in a model only dominated by this dissipative radiation component and the CC, therefore, we consider the value of $\Omega_{\Lambda}=10^{-6}$ since the contribution of radiation is bigger than the DE component in this era. Note that the fulfillment of Eq. \eqref{condicionBigRip} implies the fulfillment of Eq. \eqref{condicionLcero} which ensure a constant positive energy density.
Even more, as it was discussed in \cite{primerarticulo}, the Ricci scalar is finite with the presence of viscosity, but we recovered the divergence when we neglect the dissipation. On the other hand, from Eq. \eqref{H1deSitter} and Fig. \ref{Tm1pasado} it can be seen that, if we take the limit $\Omega_{\xi_{0}}\rightarrow 0$, then $E\rightarrow \infty$, and the solutions tend to a model with a Big-Bang singularity. Therefore, for the condition given by Eq. \eqref{condicionBigRip}, Eqs. \eqref{H1deSitter}  represent the asymptotic past behavior for the exact analytical solution \eqref{Tm1}.
 \end{itemize}

In our study, this early-time behavior of the solution is of interest, in which the model is dominated by a radiation viscous component and a DE given by the CC, which is small but not negligible, leading to a solution without the initial Big-Bang singularity, with a behavior known as soft-Big Bang, describing a solution that starts in an eternal physical past time that comes from a static universe, avoiding a quantum regime for the space-time. In particular, we study the fulfillment of the near equilibrium condition, the mathematical stability, and the entropy production of the solution in order to obtain a suitable non-singular early-time viscous solution from the physical and mathematical point of view.

\section{Near equilibrium condition, mathematical stability and entropy production}\label{tresproblemas}

\subsection{Near equilibrium condition}
The near equilibrium condition \eqref{nearequilibrium} is a consequence of the fact that Eckart's theory is a relativistic thermodynamic theory out of the equilibrium, but close to this. According to Maartens in \cite{Dissipativecosmology} and by the consideration of Eqs. \eqref{rr1}, \eqref{Peff}, and \eqref{Pi}, which leads to
\begin{equation}\label{rr}
\frac{\ddot{a}}{a}=-\frac{1}{6} \left(\rho+3\left(p+\Pi\right)\right)+\frac{\Lambda}{3},
\end{equation}
the condition to have an accelerated expansion ($\ddot{a}>0$) driven only by the negativeness of the viscous pressure $\Pi$ (i.e. $\Lambda =0$) drives to the condition
\begin{equation} \label{Pieqa}
-\Pi >p+\frac{\rho}{3}.
\end{equation}
A direct inspection of the previous result shows that the viscous pressure $\Pi$ is greater than the equilibrium pressure $p$ of the fluid, and then, the near equilibrium condition is not fulfilled. In other words, an accelerated expansion due only to the negativeness's of the viscous pressure requires a theory beyond the near equilibrium regime assumed in Eckart's theory \cite{Dissipativecosmology}. Nevertheless, according to the authors in \cite{Analysing,articulobueno,segundoarticulo}, the near equilibrium condition could hold in some regimes if one includes a positive CC. In this sense, the condition $\ddot{a}>0$ on Eq. \eqref{rr} leads to
\begin{equation}\label{Pieqa1}
-\Pi >\frac{-2\Lambda}{3}+p +\frac{\rho}{3},
\end{equation}
and the near equilibrium condition could be fulfilled because from Eq. \eqref{Pieqa1} the viscous pressure $\Pi$ is not necessarily greater than the equilibrium pressure $p$ of the dissipative fluid. On the other hand, it is possible to rewrite the near equilibrium condition \eqref{nearequilibrium} in a dimensionless form, by the use of Eq. \eqref{Pi} and the EoS, leading to
\begin{equation}\label{lhm1}
    l=\left|\frac{E(\tau)\Omega_{\xi_{0}}}{\gamma-1}\right|\ll 1.
\end{equation}    
Note that this last result shows that the behavior of $l$ as a function of time $\tau$ is driven by the behavior of the dimensionless Hubble parameter $E(\tau)$, and this condition is not satisfied when the solution has a initial singularity, where $E(\tau)\to \infty$. Nevertheless, for the early radiation dominant era, where the value of $\gamma$ is bigger than one, the last expression remains finite for a past eternal universe in which $E(\tau)\to const.$, indicating the possibility of fulfilling the near equilibrium condition. It is important to note that this result is independent of the inclusion of a CC because in a radiation domination era we need a decelerated expanding universe instead of an accelerated expanding one.

\subsection{Mathematical stability}\label{estabilidaddS}
For mathematical stability, we study the behavior of the Hubble parameter under a small perturbation, i.e., we write
\begin{equation}\label{Hperturbed}
E_{\delta}(\tau)= E(\tau)+h(\tau), \,\,\, |h(\tau)|\ll E(\tau),
\end{equation}
where $E(\tau)$ correspond to the unperturbed Hubble parameter in its dimensionless form, and $h(\tau)$ correspond to the small perturbation function. Introducing Eq. (\ref{Hperturbed}) in Eq. (\ref{Hpunto}), we obtain the following differential equation, in our dimensionless notation, for $h(\tau)$:
\begin{equation}\label{hpunto1}
h'-\frac{9\Omega_{\xi_{0}}}{2}\left(E^{2}-\frac{2\gamma}{3\Omega_{\xi_{0}}}E-\frac{\Omega_{\Lambda_{0}}}{3}\right)h=0,
\end{equation} 
where ``prime" denotes the derivative with respect to $\tau$. The above expression is a differential equation for the perturbation function $h(\tau)$ and describes the behavior of this perturbation with time $\tau$, i.e, for a mathematically stable solution to early times, we need to impose that $h(\tau\to-\infty)\to 0$.

\subsection{Entropy production}\label{seccion2A}
The internal energy of a cosmic fluid and its physical three dimensional volume are given by $U=\rho V$ and $V=V_{0}a^{3}$ (where $V_{0}$ is the present-time volume), respectively. By the consideration of the the first law of thermodynamics equation
\begin{eqnarray}\label{primeraley}
    TdS=dU+pdV, 
\end{eqnarray}
where $T$ and $S$ are the temperature and entropy of the cosmic fluid, respectively, we can find from Eq. \eqref{primeraley} the Gibbs equation \cite{Mar2,segundoarticulo} given by
\begin{equation}\label{gibbspaso1}
    dS=-\left(\frac{\rho+p}{Tn^{2}}\right)dn+\frac{d\rho}{Tn},
\end{equation}
where $n=N/V$ is the number of particle density. Also, we have the following integrability condition
\begin{equation}\label{condicionint}
    \left[\frac{\partial}{\partial \rho}\left(\frac{\partial S}{\partial n}\right)_{\rho}\right]_{n}= \left[\frac{\partial}{\partial n}\left(\frac{\partial S}{\partial \rho}\right)_{n}\right]_{\rho},
\end{equation}
that must hold on to the thermodynamical variables $\rho$ and $n$. Following this, we considered the thermodynamic assumption in which the temperature is a function of the number of particles density and the energy density, i.e., $T(n,\rho)$. With this, the integrability condition given by Eq. \eqref{condicionint} become in \cite{Tamayo, Mar2,segundoarticulo}
\begin{eqnarray}\label{temperatura}
    n\frac{\partial T}{\partial n}+\left(\rho+p\right)\frac{\partial T}{\partial \rho}=T\frac{\partial p}{\partial \rho}.
\end{eqnarray}
In order to compare our result with the model without viscosity, we study the case of a perfect fluid and a viscous fluid separately.

For a perfect fluid, the particle four-current is given by $n^{\alpha}_{;\alpha}=0$, where the symbol ``;" refers to covariant derivative, and together with the conservation equation, we have the following expressions for the particle density and the energy density, respectively:
\begin{eqnarray}\label{npunto}
    \dot{n}+3H n=\frac{\dot{N}}{N}&=&0,\\ \label{fluidoperfecto}
    \dot{\rho}+3H(\rho+p)&=&0.
\end{eqnarray}
Assuming also that the energy density depends on temperature and volume, i.e., $\rho(T,V)$  \cite{Tamayo,segundoarticulo}, we have then the following relation:
\begin{eqnarray} \label{ecuacionparaT}
\frac{d \rho}{d a}=\frac{\partial \rho}{\partial T}\frac{d T}{d a}+\frac{3n}{a}\frac{\partial \rho}{\partial n}.
\end{eqnarray}
It is possible to show, with the help of Eqs. \eqref{fluidoperfecto}, \eqref{ecuacionparaT}, and the EoS, that the temperature given by Eq. \eqref{temperatura} will be directly proportional to the internal energy, i.e., $T \sim U$ \cite{Tamayo,segundoarticulo}. Note, from Eq. \eqref{gibbspaso1}, together with Eqs. \eqref{npunto}, \eqref{fluidoperfecto}, and the EoS, that the entropy production is equal to $dS=0$, a condition which implies that there is no entropy production in the cosmic expansion, and therefore, the fluid is adiabatic.

For a viscous fluid, in the framework of Eckart's theory, the average four-velocity is chosen in a frame where there is no particle flux \cite{librocaro,Eckart}. Therefore, in this frame, the expression $n^{\alpha}_{;\alpha}=0$ is still valid, and  Eq. \eqref{npunto} doesn't change. On the other hand, from Eqs. \eqref{ConsEq1} and \eqref{Peff}, we have the following conservation equation:
\begin{equation}\label{ConsEq}
 \dot{\rho}+3H(\rho+p+\Pi)=0,
\end{equation}
which together with the Eq. \eqref{npunto} and the EoS, the Eq. \eqref{gibbspaso1} give us the following expression for entropy production \cite{Tamayo,entropiaprofe,segundoarticulo}, given in a dimensionless form:
\begin{equation}\label{entropyproduction} 
    nT\frac{d S}{d \tau}={3E^{2}\Omega_{\xi_{0}}\rho}.
\end{equation}
As we can see from the above expression, the entropy production in the viscous expanding universe is strictly positive, as long as the free parameters of the models are compatible with physical conditions like $\rho>0$, and is possible to obtain the behavior of a perfect fluid when $\Omega_{\xi_{0}}=0 $.

\section{Study of the exact solution to early-time}\label{exactearly}

In this section, we study the exact solution given by the Eq. \eqref{Tm1} under the condition \eqref{condicionBigRip} in terms of the fulfillment, at the same time, of the near equilibrium condition, the mathematical stability, and the positiveness of the entropy production. For that end, we focus our analysis on two defined early-time epochs of validity for the solution: (i) an arbitrary time for the radiation domination era in which $\tau=0$, $a=1$ and $E=1$, and (ii) the asymptotic soft-Big Bang era where $\tau\to-\infty$ and for which $E=\gamma/\Omega_{\xi_{0}}$.

\subsection{Near equilibrium condition}\label{nearequilibrionearly}
Here, we study Eq. \eqref{lhm1} backward in time. Note that this condition can be rewritten as
\begin{equation}\label{gammaWDM}
  E(\tau)\ll\frac{\gamma-1}{\Omega_{\xi_{0}}}=\frac{\gamma}{\Omega_{\xi_{0}}}-\frac{1}{\Omega_{\xi_{0}}},
\end{equation}
an expression that tells us that as long as $E(\tau)$ be much smaller than $(\gamma-1)/\Omega_{\xi_{0}}$, then the solution must be near to the thermodynamic equilibrium. Considering then the first case in which $E(\tau=0)=1$, the condition \eqref{gammaWDM} leads to
\begin{equation} \label{actualnearcondition}
    \Omega_{\xi_{0}}\ll\gamma-1,
\end{equation}
which ensures that the theory is close to the thermodynamic equilibrium in the initial radiation domination era as long as $\Omega_{\xi_{0}}\ll 1/3$. It is important to note that, if we satisfy the previous expression, then, we will satisfy, at the same time, the condition \eqref{condicionBigRip} for which this past eternal solution hold. On the other hand, as time goes backward, then the solution increase asymptotically to the de Sitter-like solution given by \eqref{H1deSitter}, and therefore, for the soft-Big Bang era the near equilibrium condition \eqref{gammaWDM} leads to
\begin{equation} \label{latenearcondition}
    \gamma\ll \gamma-1,
\end{equation}
i.e., the solution is far from the near equilibrium condition. In other words the near equilibrium condition can never be satisfied for $\gamma \geq 1/2$ (i.e., $p\geq -\rho/2$). Note that, in this case, and contrary to what happens in the case of an accelerated expanding universe, the inclusion of a positive CC constant could lead to the violation of the near equilibrium condition because a radiation domination era is characterized by a decelerated expanding universe ($\ddot{a}<0$) and, therefore, the inequality in the condition \eqref{Pieqa1} changes. Nevertheless, the not inclusion of a CC implies abandoning all the good behaviors of the solution at late times\cite{segundoarticulo}. Therefore, the soft Big-Bang solution cannot hold the near equilibrium condition in the infinity past time. In this sense, the near equilibrium condition is an effective way to study the feasibility of the solution.


In order to study the limits where the near equilibrium condition holds in the common era of radiation,
we consider in Eq. \eqref{gammaWDM} the values of $\gamma=4/3$ and $\Omega_{\Lambda_{0}}=1\times 10^{-6}$, and we assume for simplicity that
we are far from near equilibrium condition when $l=1$, i.e.,
\begin{equation}\label{Efinal}
    E=\frac{1}{3\Omega_{\xi_{0}}}.
\end{equation}
To be close to the near equilibrium condition, if we substitute the above value for the dimensionless Hubble parameter in \eqref{Tm1}, and the life-time for the radiation dominant era from Eq. \eqref{Hestandar}, we will get from Eq. \eqref{Tm1} a value for the viscosity of $\Omega_{\xi_{0}}=0.0080938$. Hence, we have the following restriction:

\begin{equation}\label{constrainnearearly}
    0<\Omega_{\xi_{0}}<0.0080938.
\end{equation}

\begin{figure}[H]
\includegraphics[width=0.5\textwidth]{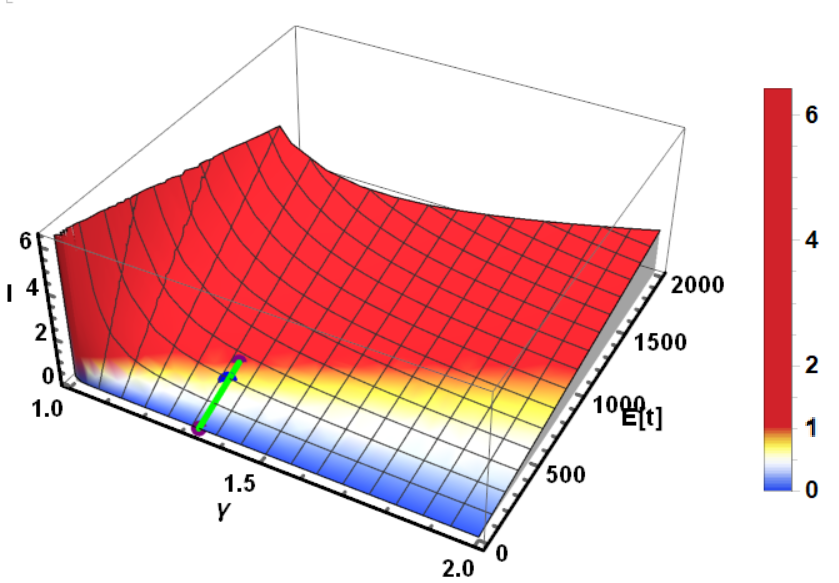}
\caption{ Behavior of $l$, given by Eq. \eqref{lhm1}, with respect to $\gamma$, and from $E=1$ (arbitrary initial radiation time) to $E=\gamma/\Omega_{\xi_{0}}$ (soft Big-Bang). The particular value of $\Omega_{\xi_{0}}=0.001$ is considered. The green line represents the near equilibrium history when $\gamma=4/3$, and the blue point represents the moment when the Big Bang occurs for the $\Lambda$CDM model.}\label{lparaearly}
\end{figure}

In Fig. \ref{lparaearly} we display the near equilibrium condition for the fixed value of $\Omega_{\xi_{0}}=0.001$. At the beginning of the radiation domination era $(\tau = 0$ and $E=1$), we obtained a value of  $l=0.003$, according to the Eq. \eqref{lhm1}, and we are far from near equilibrium, according to Eq. \eqref{Efinal}, when $E=333.333$. The purple points represent the maximum and minimum values of $l$ in order to be close to the near equilibrium when $\gamma=4/3$, and the green line represents the evolutionary history in which we are near to thermodynamic equilibrium. The blue point represents the value when $E=235.704$, which represents the point where the Big Bang should occur if we consider the lifetime of the Universe according to Eq. \eqref{Hestandar} for the $\Lambda$CDM model in Eq. \eqref{Tm1}.

Therefore, the solution can describe the radiation domination era of the $\Lambda$CDM model with a viscous radiation component that fulfills, for the entire era, the near equilibrium condition, but, for the infinite past time, when the solution tends to the soft Big-bang, we need to postulate that the solution holds beyond the near equilibrium condition.

\subsection{Mathematical stability}\label{estabilidadearlyexacta}
To analyze the mathematical stability, we use Eq. \eqref{hpunto1}, changing the integration variable from $\tau$ to $E$, obtaining in our dimensional notation the expression
\begin{equation}
  \frac{dh}{dE}=\frac{\left(\Omega_{\Lambda_{0}}\Omega_{\xi_{0}}+2\gamma E -3E^{2}\Omega_{\xi_{0}}\right)h}{\left(E^{2}-\Omega_{\Lambda_{0}}\right)\left(\gamma-E\Omega_{\xi_{0}}\right)},
\end{equation}
whose integration leads to
\begin{equation}\label{henfunciondeE}
    h(E)=h_{i}\left(E\Omega_{\xi_{0}}-\gamma\right)\left(\Omega_{\Lambda_{0}}-E^{2}\right),
\end{equation}
being $h_{i}$ and integration constant. If the condition given by Eq. \eqref{condicionBigRip} holds, then $E$ tends to $\gamma/\Omega_{\xi}$ backward in time (soft Big-Bang), and the perturbation function given by Eq. \eqref{henfunciondeE} goes to zero at this time, i.e, the solution is mathematically stable when $\tau\rightarrow-\infty$ or, in other words, the soft Big-Bang solution is mathematically stable. It is important to note that the solution \eqref{Tm1} is also mathematically stable when $\tau\rightarrow\infty$ in which $E=\sqrt{\Omega_{\Lambda_0}}$ (de Sitter solution), as can be seen from Eq. \eqref{henfunciondeE}. 

\begin{figure}[H]
\centering
\includegraphics[width=0.5\textwidth]{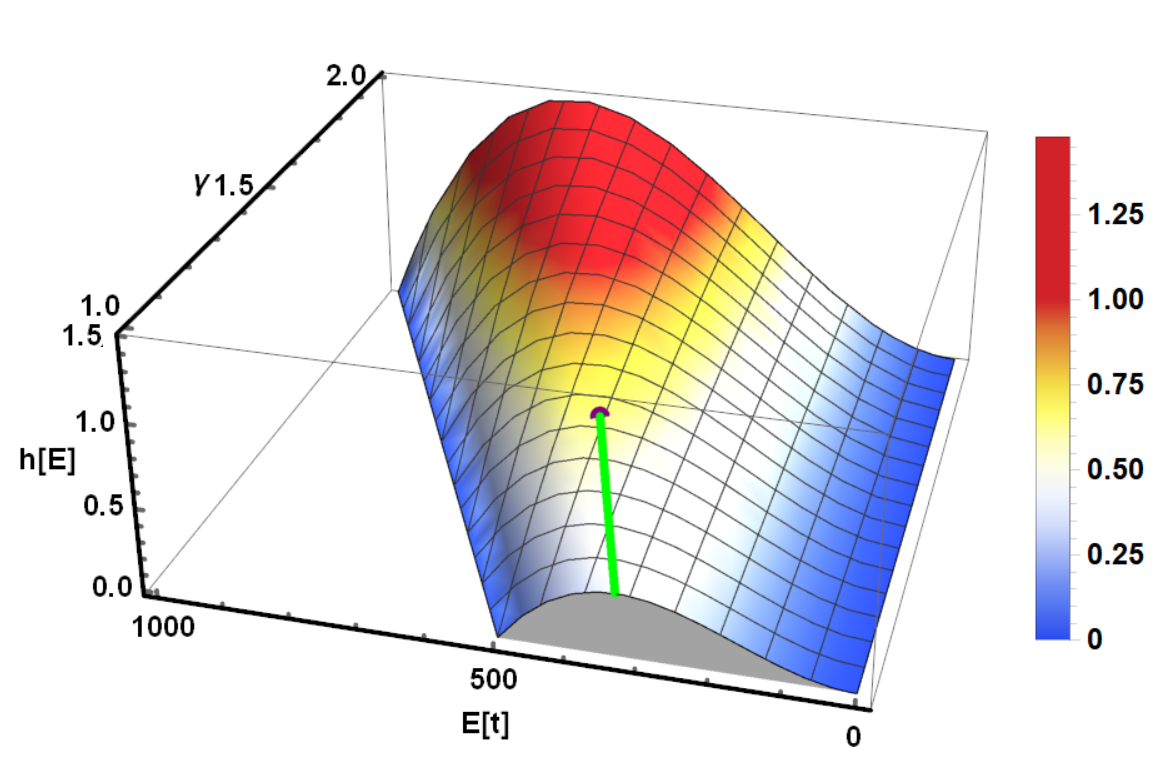}
\caption{Behavior for $h$, given by Eq. \eqref{henfunciondeE}, as a function of $\gamma$, and from $E=1$ (arbitrary initial radiation-time) to $E=\gamma/\Omega_{\xi}$ (soft Big-Bang). The values of $\Omega_{\Lambda}=10^{-6}$ and $\Omega_{\xi}=0.002$ are considered. The green line are given by Eq. \eqref{condicionparametros}, and the purple point represent the maximum value obtained for $\gamma=4/3$.}
\label{conclusionfinal}
\end{figure}

We can find the maximum values for the perturbation function form Eq. \eqref{henfunciondeE}, being obtained
\begin{equation}\label{puntoscriticos}
    E=\frac{\gamma\pm\sqrt{\gamma^2+3\Omega_{\xi_{0}}\Omega_{\Lambda_{0}}}}{3\Omega_{\xi_{0}}}.
\end{equation}
From now, we only consider the positive sign, because the negative sign gives a value of $E<0$ which doesn't occur for the entire cosmic evolution of the solution \eqref{Tm1}.
In order to find some constraint in the cosmological parameters that ensure stability in the entire radiation era, we substitute Eq. \eqref{puntoscriticos} in \eqref{henfunciondeE}, getting the following condition:
\begin{eqnarray}\label{hmaximo}
\nonumber
    h&=&  h_{i}\Big|\Big(\frac{2 \gamma^3}{27 \Omega^{2}_{\xi_{0}}}+\frac{2 \gamma ^2 \sqrt{\gamma ^2+3 \Omega^{2}_{\xi_{0}} \Omega_{\Lambda_{0}} }}{27 \Omega^{2}_{\xi_{0}}}\\ \label{condicionparametros}
    &+&\frac{2}{9} \Omega_{\Lambda_{0}} \sqrt{\gamma ^2+3 \Omega^{2}_{\xi_{0}} \Omega_{\Lambda_{0}} }-\frac{2 \gamma  \Omega_{\Lambda_{0}} }{3}\Big)\Big|\ll E(\tau),
\end{eqnarray}
where  $h_{i}$ is fixed considering that $h(E=1)=0.00001$. From Eq. \eqref{hmaximo}, we can see that if $\Omega_{\xi_{0}}\rightarrow 0$, then $h\rightarrow\infty$ and we are far from the mathematical stability. This should not be a surprise because with very small values of $\Omega_{\xi_{0}}$ the solution tends to the $\Lambda$CDM model with Big-Bang singularity as was discussed in section \ref{seccion2}. For the radiation era, $\Omega_{\Lambda_{0}}$ has a small contribution to the total energy budget and, in addition, in order to be close to the near equilibrium condition, we need to fulfill the constraint given by Eq. \eqref{constrainnearearly}, i.e., $\Omega_{\xi_{0}}$ must be small. We begin studying the stability when $H(t)$ is close to an arbitrary value of $H_{0}$, in the radiation dominant era. Therefore, we can approximate Eq. \eqref{hmaximo} as

\begin{equation}\label{hmaximo2}
    h\approx\left|h_{i} \frac{4\gamma^3}{27 \Omega^{2}_{\xi_{0}}}\right|<1.
\end{equation}
For the radiation era, we have $\gamma=4/3$, and the above restriction in $\Omega_{\xi_{0}}$ gives us
\begin{equation}\label{restriccionxiradiacion}
    0.00162387<\Omega_{\xi_{0}}<\frac{4}{3}.
\end{equation}
In Fig. \ref{conclusionfinal}, we display the behaviour of the perturbation function $h$ as a function of the dimensionless Hubble parameter $E$, given by Eq. \eqref{henfunciondeE}, for the fixed values of $\Omega_{\xi_{0}}=0.002$ and $\Omega_{\Lambda}=10^{-6}$. From Eq. \eqref{hmaximo2}, we obtain that the maximum value for $h$ is $0.659426$, and from Eq. \eqref{puntoscriticos}, we can see that this value is obtained when $E=444.444$. Therefore,
the solution \eqref{Tm1} is stable for the usual radiation domination era. In the figure, the green line is given by Eq. \eqref{condicionparametros} and represents the maximum points that can be obtained for the value of  $h(E)$. We can also see one point in purple that represents the maximum value obtained for radiation ($\gamma=4/3$).


Finally, we can find a single constraint on $\Omega_{\xi_{0}}$ from Eqs. \eqref{constrainnearearly} and \eqref{restriccionxiradiacion}, given by 

\begin{equation}\label{restriccioncombinada}
  0.00162387< \Omega_{\xi_{0}} <0.0080938.
\end{equation}
This single constraint allows us to satisfy both criteria,  the near equilibrium condition, and the mathematical stability in the Hubble parameter, for the usual radiation era of the $\Lambda$CDM model.




\subsection{Temperature and entropy production}\label{seccion5C}
In order to evaluate the entropy production of the dissipative radiation component of the model from the Eq. \eqref{entropyproduction}, we first need to find their temperature from the Eq. \eqref{temperatura}. Rewriting the conservation Eq. \eqref{ConsEq} in the form
\begin{equation}\label{pararhoa}
    \frac{d\rho}{da}=-\frac{3\rho}{a}\left(\gamma-3H\xi_{0}\right),
\end{equation}
we can rewrite Eq. \eqref{ecuacionparaT} as
\begin{equation}\label{paso1}
    \rho\left(\gamma-3H\xi_{0}\right)=-\frac{a}{3}\frac{\partial \rho}{\partial T}\frac{dT}{da}-n\frac{\partial \rho}{\partial n}.
\end{equation}
Then, from Eq. \eqref{temperatura} and Eq. \eqref{Peff}, we have
\begin{equation}\label{paso2}
    n\frac{\partial T}{\partial n}+\rho \left(\gamma-3H\xi_{0}\right)\frac{\partial T}{\partial \rho}=T\left[\left(\gamma-1\right)-3H\xi_{0}-3\xi_{0}\rho\frac{\partial H}{\partial \rho}\right],
\end{equation}
which together with Eq. \eqref{paso1} leads to
\begin{equation}\label{paso8}
\frac{dT}{T}=-3\frac{da}{a}\left[\left(\gamma-3H\xi_{0}\right)-1-3\xi_{0}\rho\frac{\partial H}{\partial \rho}\right].   
\end{equation}
By the use of Eqs. \eqref{tt} and \eqref{pararhoa}, we can rewrite the Eq. \eqref{paso8} in the the following form:
 \begin{equation}\label{diferentialT}
\frac{d T}{T}=\frac{d\rho}{\rho}\left[1-\frac{\left(\frac{2}{3}\sqrt{3\left(\rho+\Lambda\right)}+\xi_{0}\rho\right)}{\frac{2}{3}\sqrt{3\left(\rho+\Lambda\right)}\left(\gamma-\sqrt{3\left(\rho+\Lambda\right)}\xi_{0}\right)}\right].
 \end{equation}
Since we are considering the evolution of the solution \eqref{Tm1} in an arbitrary time in the radiation domination era in which $E(\tau=0)=1$ to the past when $E(\tau\to -\infty)=\gamma/\Omega_{\xi_{0}}$, then, we integrate the Eq. \eqref{diferentialT} from $E=1$ to an arbitrary past-time $E(\tau<0)$, being obtained
\begin{eqnarray}
&\ln{\left(\frac{T}{T_{0}}\right)}=\ln\left(\frac{\rho}{\rho_{0}}\right)\nonumber\\
&-\frac{2\Omega_{\xi_{0}}\sqrt{\Omega_{\Lambda}}\left[\arctanh{\left(\frac{\sqrt{\Omega_{\Lambda_{0}}}}{E}\right)}-\arctanh{\left(\sqrt{\Omega_{\Lambda_{0}}}\right)}\right]}{\left(\gamma^{2}-\Omega_{\Lambda_{0}}\Omega^{2}_{\xi_{0}}\right)}+\nonumber\\ \label{paso3}
&\frac{\gamma\ln\left({\frac{\rho}{\rho_{0}}}\right)-\left[\gamma(2+\gamma)-\Omega_{\Lambda_{0}}\Omega^{2}_{\xi_{0}}\right]\ln{\left(\frac{\gamma-E\Omega_{\xi_{0}}}{\gamma-\Omega_{\xi_{0}}}\right)}}{\left(\gamma^{2}-\Omega_{\Lambda_{0}}\Omega^{2}_{\xi_{0}}\right)}.
\end{eqnarray}

On the other hand, integrating Eq. \eqref{pararhoa}, with the help of Eq. \eqref{tt}, we obtain in our dimensionless notation the expression
\begin{eqnarray}\label{rhoenfunciondea}
\ln{a^{3}}&=&\frac{2\Omega_{\xi_{0}}\sqrt{\Omega_{\Lambda}}\left[\arctanh{\left(\frac{\sqrt{\Omega_{\Lambda_{0}}}}{E}\right)}-\arctanh{\left(\sqrt{\Omega_{\Lambda_{0}}}\right)}\right]}{\left(\gamma^{2}-\Omega_{\Lambda_{0}}\Omega^{2}_{\xi_{0}}\right)}\nonumber\\ \label{paso4}
&+&\frac{-\gamma\ln\left({\frac{\rho}{\rho_{0}}}\right)+2\gamma\ln{\left(\frac{\gamma-E\Omega_{\xi_{0}}}{\gamma-\Omega_{\xi_{0}}}\right)}}{\left(\gamma^{2}-\Omega_{\Lambda_{0}}\Omega^{2}_{\xi_{0}}\right)}.
\end{eqnarray}
With this last result, we can express the temperature from Eq. \eqref{paso3} of the dissipative radiation fluid as a function of the scale factor as follows
\begin{equation}\label{temperaturaearly}
    T=T_{0}\left(\frac{\rho}{\rho_{0}}\right)\left(\frac{\gamma-\Omega_{\xi_{0}}}{\gamma-E\Omega_{\xi_{0}}}\right){a^{-3}}.
\end{equation}

We can see from Eq. \eqref{temperaturaearly} that when $E(\tau\rightarrow-\infty)$ the solution goes to the soft Big-Bang given by Eq. \eqref{H1deSitter}, and the temperature increase as long as the radiation fluid is compressed in the past infinity ($a\rightarrow0$), until reaching a state where the energy density would be constant and given by Eq. \eqref{rhodeSitter1}. In this stage, the universe would be infinitely hotter since the expression for the temperature of the dissipative fluid given by Eq. \eqref{temperaturaearly} diverge, being this a reasonable behavior for which the solution would be far from the near equilibrium regime, as it was discussed in the section \ref{nearequilibrionearly}. It is important to note that, in this model, the temperature increases to infinity in the infinity past, whereas in the $\Lambda$CDM model the temperature is infinity in a finite pats time.

\begin{figure}[H]
\centering
\includegraphics[width=0.5\textwidth]{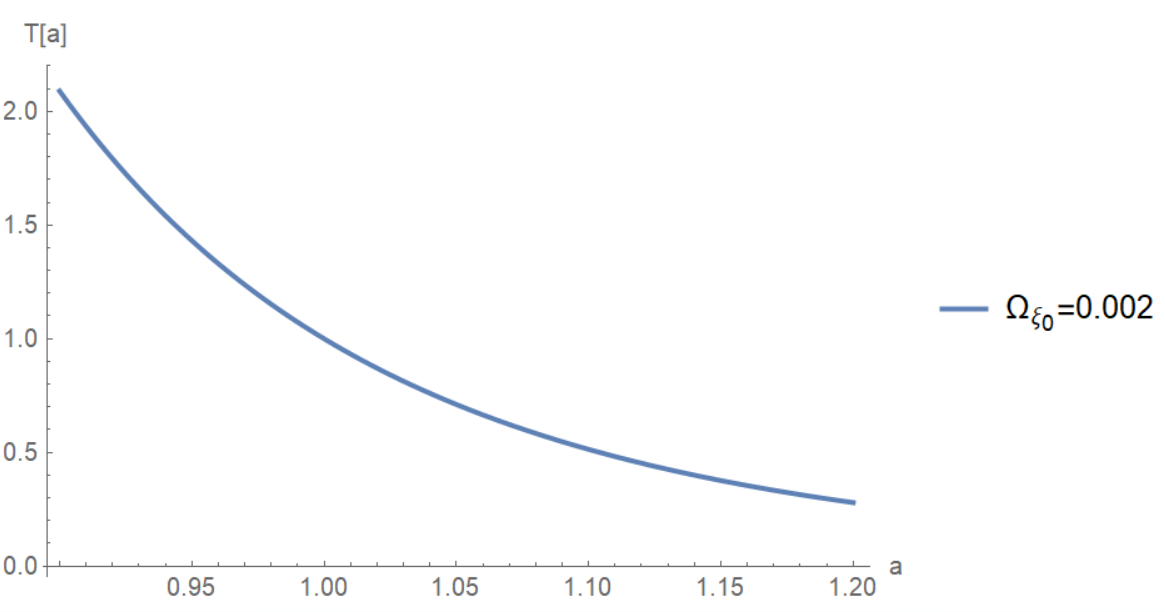}
\caption{Behavior of the temperature of the dissipative radiation fluid ($\gamma=4/3$), given by Eq. (49), from $a=0.9$ to $a=1.20$ ($a=1$ represents an arbitrary time for the radiation domination era in which $\tau=0$ and $E=1$). The values of $\Omega_{\Lambda_{0}}=10^{-6}$ and $\Omega_{\xi_{0}}=0.002$ are considered.}
\label{figuraTearly}
\end{figure}

In Fig. \ref{figuraTearly}, we display the numerical behavior of the temperature for the dissipative radiation fluid  ($\gamma=4/3$) for the fixed values of $\Omega_{\Lambda_{0}}=10^{-6}$ and $\Omega_{\xi_{0}}=0.002$ (according to restriction \eqref{restriccioncombinada}). The energy density of the dissipative radiation fluid is obtained from Eq. \eqref{rhoenfunciondea}.

With the temperature given by Eq. \eqref{temperaturaearly}, we can calculate the entropy production from Eq. \eqref{entropyproduction}. By the use of Eq. \eqref{npunto} we have $n=n_{0}a^{3}$, and the entropy production for the dissipative radiation fluid is given by
\begin{equation}\label{entropyfinal??}
  \frac{dS}{d\tau}=\frac{3E^{2}\Omega_{\xi_{0}}\rho}{nT}=\frac{3E^{2}\rho_{0}\left(\gamma-E\Omega_{\xi_{0}}\right)\Omega_{\xi_{0}}}{n_{0}T_{0}\left(\gamma-\Omega_{\xi_{0}}\right)}.
\end{equation}
Note that, in the infinite past, the solution \eqref{Tm1} tends to the asymptotic soft Big-bang solution $E=\gamma/\Omega_{\xi_{0}}$ with null entropy production according to Eq. \eqref{entropyfinal??}. This result is due to the asymptotic behavior of the fluid since, in this stage, the fluid will be fully compressed with a constant energy density given by Eq. \eqref{rhodeSitter1} with no change in the micro-states of the system. Considering Eq. \eqref{restriccioncombinada}, then the entropy production for this radiation domination era ($\gamma=4/3$) would be positive and finite. 

On the other hand, making use of Eq. \eqref{Hpunto} to rewrite Eq. \eqref{entropyfinal??}, and making the integration over $E$, we will obtain in our dimensional notation the expresion
\begin{equation}\label{dSEearly}
dS=\frac{-2\Omega_{\xi_{0}}\rho_{0}E^{2}dE}{n_{0}T_{0}\left(\gamma-\Omega_{\xi_{0}}\right)\left(E^{2}-\Omega_{\Lambda_{0}}\right)},   
\end{equation}
and then,
\begin{eqnarray}\label{entropiafinalearly}
     S(E)&=&S_{0}-\frac{2 \rho_{0}}{n_{0}T_{0}\left(\gamma -\Omega_{\xi_{0}}\right) }\Big\{\nonumber \\ &+&\Omega_{\xi_{0}}\sqrt{\Omega_{\Lambda_{0}}} \ln\left[\left(\frac{\sqrt{\Omega_{\Lambda_{0}}}\Omega_{\xi_{0}}+\gamma}{\sqrt{\Omega_{\Lambda_{0}}}\Omega_{\xi_{0}}-\gamma}\right)\left(\frac{\sqrt{\Omega_{\Lambda_{0}}}-E}{\sqrt{\Omega_{\Lambda_{0}}}+E}\right)\right]\nonumber \\
     &+&E\Omega_{\xi_{0}}-{\gamma }\Big\}.
 \end{eqnarray}
 
\begin{figure}[H]
\includegraphics[width=0.5\textwidth]{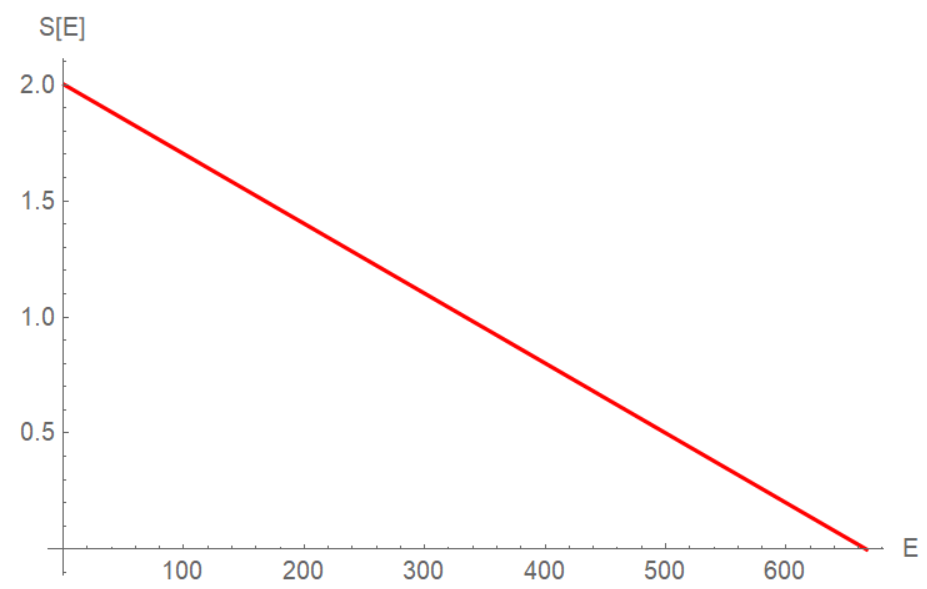}
\caption{Behavior of the entropy for the dissipative radiation fluid ($\gamma=4/3$), given by Eq. \eqref{entropiafinalearly}, from $E=1$ (arbitrary initial time in the radiation domination era) to $E=\gamma/\Omega_{\xi_{0}}$ (soft Big-Bang). The values of $\Omega_{\Lambda_{0}}=10^{-6}$ and $\Omega_{\xi_{0}}=0.002$ are considered.}
\label{entropiaearly}
\end{figure}

From the above expression, the constraint for a positive energy density for the soft Big-bang solution given by Eq. \eqref{condicionLcero} ensures a real value for entropy since in the logarithm term $E$ is greater than $\sqrt{\Omega_{\Lambda_{0}}}$. Note that, for the infinity past when $E\rightarrow\gamma/\Omega_{\xi_{0}}$, we get $S=S_{0}$, i.e. a constant entropy. In Fig. \ref{entropiaearly} is represented the numerical behavior of the entropy for the dissipative radiation fluid ($\gamma=4/3$) for $\Omega_{\Lambda_{0}}=10^{-6}$ and $\Omega_{\xi_{0}}=0.002$ (according to restriction Eq. \eqref{restriccioncombinada}), for the particular initial value of $S_{0}=0$. We can see that the same condition that satisfies the near equilibrium condition and the mathematical stability of the Hubble parameter are compatible with a positive entropy that increase with the universe expansion from $E=\gamma/\Omega_{\xi_{0}}=666.7$ (soft-Big Bang) to $E=1$ (arbitrary initial time in the radiation domination era).

\section{Discussion}\label{seccionfinal}
We have studied throughout this work the near equilibrium condition, the mathematical stability of the Hubble parameter, and the entropy production of a bulk viscous model previously found in \cite{primerarticulo}, within the framework of Eckart's theory, in an early-time era of the cosmic evolution. The model represents a universe filled with a dissipative radiation component, where the bulk viscosity is proportional to their energy density according to the expression $\xi=\xi_{0}\rho$, and a positive CC whose contribution to the total energy density budget is small but not negligible. This model is characterized by a past eternal behavior known as ``soft-Big Bang", with an asymptotic solution given by Eq. \eqref{H1deSitter}, as long as the condition $\Omega_{\xi_{0}}<\gamma$ holds, which also implies the fulfillment of the condition \eqref{condicionLcero}, i.e., this asymptotic behavior has a constant and positive energy density in the infinity past time, contrary to the infinity energy density obtained in the $\Lambda$CDM model, which occurs in a finite time. In addition, this model follows very close to the $\Lambda$CDM behavior for small values of dissipation ($\Omega_{\xi}\rightarrow0$), as we can see in Fig. \ref{Tm1pasado}. 

We have shown in section \ref{nearequilibrionearly} that, for a viscous radiation component, where $\gamma$ is bigger than one, the behavior of the expression for the near equilibrium condition ($l$) is driven by the behavior of the dimensionless Hubble parameter $E(\tau)$, according to Eq. \eqref{lhm1}, and, as a long as the condition \eqref{condicionBigRip} is satisfied, $E(\tau)$ tends asymptotically to the soft-Big bang solution \eqref{H1deSitter} and, therefore, the near equilibrium condition \eqref{lhm1} remains finite for this past eternal universe. From Fig. \ref{lparaearly}, we can see that this model can describe the radiation domination era of the $\Lambda$CDM model and, for the entire era, the near equilibrium condition is fulfilled as a long as the constraint for the bulk viscous coefficient given by Eq. \eqref{constrainnearearly} is satisfied. Nevertheless, for the infinite past time, when the solution tends to the soft-Big bang, the near equilibrium condition is not satisfy. Thus, we examine the limits within which the solution remains viable in terms of the near equilibrium condition.


In section \ref{estabilidadearlyexacta}, we have shown that a small perturbative function in the mathematical expression for the Hubble parameter solution \eqref{Tm1} is stable for $\gamma=4/3$, as long as we satisfy the condition given by Eq. \eqref{restriccionxiradiacion}, as we can see in Fig. \ref{conclusionfinal}. Therefore, we can have a single constraint in $\Omega_{\xi_{0}}$ to satisfy both criteria, the near equilibrium condition and the stability in the Hubble parameter which is given by Eq. \eqref{restriccioncombinada}.

The temperature and the entropy production for the viscous radiation fluid were studied in section \ref{seccion5C}. The temperature is given by  Eq. \eqref{temperaturaearly}, and in the infinity past ($a\rightarrow0$), the fluid is compressed to a constant energy density given by Eq. \eqref{rhodeSitter1}, where the temperature diverges and starts to decrease during his expansion as we can see in Fig. \ref{figuraTearly}. Therefore, as long as the temperature increase to infinity, we will start to be far from near equilibrium condition as is displayed in Fig. \ref{lparaearly}. The advantage of this model is that the temperature is not infinity in a finite time like in the $\Lambda$CDM model and, therefore, the presence of viscosity together with a positive CC acts as a modulator in the near equilibrium process before the temperature starts to be extremely high. For the entropy production, this is strictly positive, and when we go backward in time the fluid starts to compress as long as the scale factor goes to zero, getting a constant energy density given by Eq. \eqref{rhodeSitter1}. Therefore, the entropy production goes to zero during this compression of the fluid, and the restriction \eqref{condicionLcero} appears automatically within the expression of the logarithm in \eqref{entropiafinalearly}, to ensure the real value of the entropy. Also, the consideration of near equilibrium regime together with mathematical stability implies the condition \eqref{restriccioncombinada}, and with this constraint, the entropy is positive and well behaved, as we can see in Fig. \ref{entropiaearly}.

Therefore, the solution given by Eq. \eqref{Tm1} can describe the entire radiation era of the $\Lambda$CDM model with a viscous radiation component instead of a perfect radiation fluid, satisfying at the same time the near equilibrium condition, the positiveness of the entropy production and the solution is mathematically stable, as a long as the conditions described before holds. The principal characteristic of the solution is to avoid the initial Big Bang singularity, although the solution behaves very similar to the $\Lambda$CDM model, in an asymptotic past eternal behavior known as soft-Big Bang.
\section*{Acknowledgments}
Norman Cruz acknowledges the support of Universidad de Santiago de Chile (USACH), through Proyecto DICYT N$^{\circ}$ 042131CM, Vicerrector\'ia de Investigaci\'on, Desarrollo e Innovaci\'on. Esteban Gonz\'alez thanks to Vicerrector\'ia de Investigaci\'on y Desarrollo Tecnol\'ogico (VRIDT) at Universidad Cat\'olica del Norte (UCN) by the scientific support of N\'ucleo de Investigaci\'on No. 7 UCN-VRIDT 076/2020, N\'ucleo de Modelaci\'on y Simulaci\'on Cient\'ifica (NMSC). Jose Jovel acknowledges ANID-PFCHA/Doctorado Nacional/2018-21181327.

\bibliography{bibliografia.bib}
\end{document}